\documentclass[aps,prl,twocolumn,showpacs]{revtex4}
\usepackage{epsfig}
\usepackage{graphicx}

\begin{document}

\DeclareGraphicsExtensions{.eps,.EPS}

\title{A Feshbach resonance in d-wave collisions}
\author{Q. Beaufils, A. Crubellier$^{\diamond}$, T. Zanon, B. Laburthe-Tolra, E. Mar\'echal, L. Vernac, and O. Gorceix}
\affiliation{Laboratoire de Physique des Lasers, CNRS UMR 7538, Universit\'e Paris 13,
99 Avenue J.-B. Cl\'ement, 93430 Villetaneuse, France}
\affiliation{$^{\diamond }$ Laboratoire Aim\'e Cotton, CNRS II, B\^atiment 505, Campus d'Orsay, 91405 Orsay Cedex, France }

\begin{abstract}
We analyse a narrow Feshbach resonance with ultra-cold chromium atoms colliding in d-wave. The resonance is made possible by dipole-dipole interactions, which couple an incoming $l=2$ collision channel with a bound molecular state with $l=0$. We find that three-body losses associated to this resonance increase with temperature, and that the loss feature width as a function of magnetic field also increases linearly with temperature.  The analysis of our experimental data shows that the Feshbach coupling is small compared both to the temperature and to the density limited lifetime of the resonant bound molecular state. One consequence is that the three body losse rate is proportionnal to the square of the number of atoms, and that we can directly relate the amplitude of the losses to the Feshbach coupling parameter. We compare our measurement to a calculation of the coupling between the collisionnal channel and the molecular bound state by dipole-dipole interactions, and find a good agreement, with no adjustable parameter. An analysis of the loss lineshape is also performed, which enables to precisely measure the position of the resonance. 
\end{abstract}


\date{\today}

\maketitle

\section{I Introduction}

The study of atomic collisional properties in ultracold gases has been made possible by the recent development of laser cooling and trapping techniques. In a Feshbach resonance the scattering cross section is resonantly enhanced when the energy of a bound state in a given molecular potential matches the molecular dissociation limit of another molecular potential, provided both potentials are coupled. For polarized bosons at low temperatures, collision processes are dominated by s-wave scattering, and a Feshbach resonance results in a modification of the s-wave scattering length.  This feature is particularly relevant for Bose-Einstein condensates, since it allows for the production of BEC with tunable interactions, with for example possible application to interferometric measurements \cite{tunable}. Feshbach resonances can also be used to produce molecules, by ramping the magnetic field through the resonance \cite{Kohler}. In the case of mixtures of fermions, s-wave Feshbach resonances are extensively used to study the BEC-BCS crossover around the unitary limit \cite{bloch}. p-wave Feshbach resonances were also studied in the case of polarized fermions \cite{regal}. In that case, both the incoming open channel and the molecular bound state have a p-wave character. A d-wave shape resonance was also studied by colliding two Rb condensates at large velocities \cite{buggle}.  Resonances with incoming $l$ larger than zero may offer a means to study the effect of anisotropic interactions in quantum degenerate gases. 

Here, we study the resonant coupling of an incoming channel $l=2$ to a bound state of partial wave $l=0$. The coupling is due to dipole-dipole interactions. As the bound state is localized at relatively short internuclear distance (typically 100 $a_0$), the particles in the incoming channel need to tunnel through the centrifugal barrier associated to $l=2$. We shall see that tunnelling through this barrier has profound consequences on the temperature dependence of the resonance width, on the temperature dependence of the three-body loss parameter, and even on the dynamics of atom losses. 

The paper is organized as follows. We first report on an experimental study of a Feshbach resonance in d-wave collisions between ultra cold  $^{52}Cr$ atoms. We will refer to such resonance as a 'd-wave Feshbach resonance'. Losses are analysed, and we show that three-body processes lead to losses proportional to the square of the number of atoms, $N^2$, not to $N^3$ as is usually the case. We study the temperature dependence of both the amplitude of the loss feature and of its width as a function of the magnetic field $B$. Both vary linearly with $T$. We then develop a theoretical model taking into account that the Feshbach coupling is small compared to the inverse (density limited) lifetime of the bound state. We finally compare the loss amplitude with a calculation taking into account the coupling of the incoming d-wave channel to the s-wave molecular state by dipole-dipole interactions. We find a very good agreement between our data and this calculation involving no adjustable parameter.

\section{II Experimental results}

One manifestation of the strength of dipole-dipole interactions in chromium is the existence of d-wave Feshbach resonances. In the one we study, first reported in \cite{werner}, the incoming channel $\left|S=6,M_{S}=-6,l=2,m_{l}=1\right\rangle$ is coupled to $\left|S=6,M_{S}=-5,l=0,m_l=0\right\rangle$ through dipole-dipole interactions. The d-wave scattering cross section, usually vanishingly small at low temperature because the atoms need to tunnel through a centrifugal barrier, is resonantly enhanced when the energy of the last vibrational bound state of $\left|S=6,M_{S}=-5,l=0,m_l=0\right\rangle$ matches the molecular dissociation limit of $\left|S=6,M_{S}=-6,l=2,m_{l}=1\right\rangle$.

\begin{figure}
\centering
\includegraphics[width= 2.8 in]{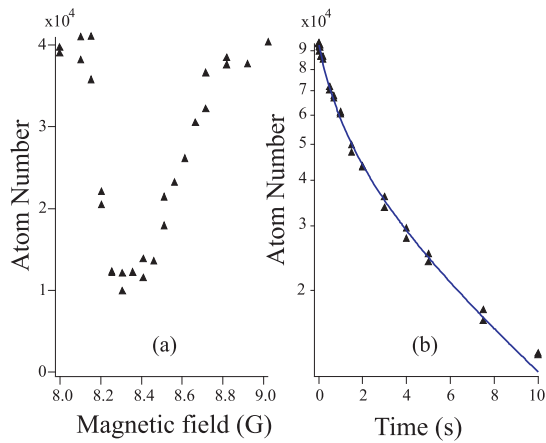}	
\caption{\setlength{\baselineskip}{6pt} {\protect\scriptsize
Left: number of atoms remaining in the trap after a hold time of 10 s, in the vicinity of the Feshbach resonance, at temperature 15 $\mu$K. Right: typical time evolution of the number of atoms after the magnetic field is set at the magnetic field corresponding to maximum losses. The sold line is  the result of a fit using eq. (\ref{Nt}).}} \label{figure1}
\end{figure}

We proceed by measuring the atom loss rate in an optically trapped Chromium atom cloud as a function of the magnetic field $B$. We first load from a MOT up to 5 million ground state $^{52}Cr$ atoms at a temperature of 100 $\mu$K in an optical dipole trap formed by a retroreflected far red detuned $35$ W laser beam, following an experimental procedure described in \cite{sweeprf}. To prevent inelastic collisions due to dipolar relaxation, we polarize the atoms in the lowest energy Zeeman state $m_{S}=-3$, which is done by optically pumping the atoms with a  circularly polarized laser at 427 nm ($i.e.$ resonant with the $^7S_3 \longrightarrow ^7P_3$ transition) in a $2.3$ G homogeneous magnetic field. The next step is to form a dimple by adding a strong longitudinal confinement to the trap. For this we transfer power from the horizontal beam to a crossed vertical beam by turning a half wave plate on a computer controlled rotation stage, in front of a polarizing beam splitter cube (as in \cite{beaufilsBEC}). During the linear rotation of this plate the depth of the horizontal beam trap decreases, forcing evaporation and thus decrasing the number of atoms and the temperature of the cloud. Determined by the rotation angle of the half-wave plate, the atom number is between $10^5$ and $ 3 \times 10^4$ and the temperature between 16 $\mu$K and 2 $\mu$K. At this point we tune the static homogeneous magnetic field to look at field dependent losses due to the d-wave Feshbach resonance around $B=$ 8.2 G. We use RF spectroscopy to calibrate the magnetic field with an uncertainty of 2 mG, and resonant absorption imaging to extract the temperature and the atom number.

Fig \ref{figure1} (a) shows the number of atoms remaining in the trap after $10$ s as a function of $B$, at $T =$ 15 $\mu$K. The lineshape of the loss feature is clearly asymmetric, which is in contrast to what is observed in most s-wave Feshbach resonances analyzed to date, but analogous to what is reported in the case of p-wave Feshbach resonances \cite{regal}. This paper provides a full experimental and theoretical analysis of these losses.

\begin{figure}
\centering
\includegraphics[width= 2.8 in]{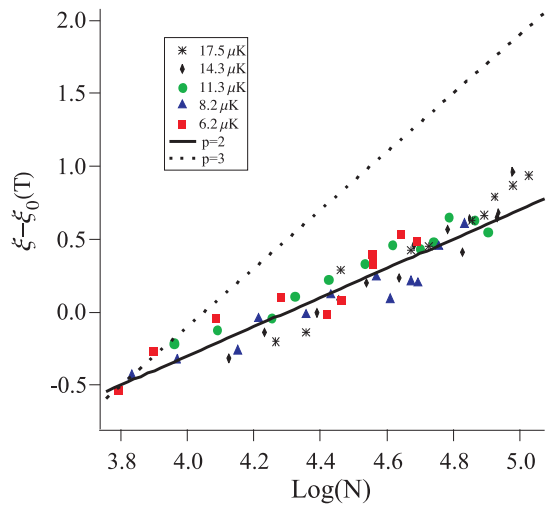}	
\caption{\setlength{\baselineskip}{6pt} {\protect\scriptsize
$\xi$ (see definition in text) measured as a function of the logarithm of the number of atoms. To better emphasize the power law dependence in $Log(N)$, the different values of $\xi$ obtained at different temperatures have been shifted so that they coincide for a given atom number. Then all values align on a linear curve of slope $p-1=1$ (solid line; the dotted line is a linear curve of slope $p=3$).}} \label{figure2body3body}
\end{figure}

As a first step to characterize the losses, we measure the time evolution of the number of atoms after the magnetic field is set at the value corresponding to the peak losses. As shown in Fig \ref{figure1}b, we observe a non-exponential decay of the number of atoms as a function of time. As the temperature of the cloud does not change much during the decay (less than 15 percent), at any given moment, the density is approximately proportional to the number of atoms; a non-exponential atom number decay is then clearly related to a density-dependent loss mechanism. We describe the density dependent losses by the following rate equation:

\begin{equation}
\frac{d n}{d t}= -\Gamma_{1} n - K_{p} n^{p}
\label{dndt}
\end{equation}
where $n$ is the atom density. $\Gamma_{1}$ is the one-body loss rate due to backgound gas collisions, and $K_{p}$ represents the density dependent loss rate parameter. Although atoms are in the lowest state of energy, so that two-body losses are energetically forbidden, and three-body losses are the most likely, we \textit{do not} assume that $p=3$. When integrated over space assuming a thermal gaussian distribution, Eq. (\ref{dndt}) leads to 

\begin{equation}
\frac{1}{N} \frac{d N}{d t}+\Gamma_{1}= - \beta_{p} N^{p-1}
\label{intdndt}
\end{equation}
where $\beta_{p}= \frac{K_p}{p^{3/2}(x_0y_0z_0 \pi^{3/2})^{p-1}}$, with $(x_0,y_0,z_0)$ the $1/e$ radii of the trapped gas. It is therefore interesting to plot $\xi \equiv Log \left(\frac{1}{N} \frac{d N}{d t}+\Gamma_{1} \right)$ as a function of $Log\left(N\right)$, as the slope gives a measure of $p-1$. Such a plot is represented in Fig \ref{figure2body3body}. $\Gamma_1 = 1/20 $s$^{-1}$ is deduced from the time constant of decay when the number of atoms is very small. For this figure, we used data obtained at five different temperatures. As we are only interested in the power number dependence $(p-1)$ of the loss process, we have shifted the experimental curves vertically by a value $\xi_0 (T)$, so that all values of $\xi - \xi_0 (T)$ coincide for a given atom number. It can be seen in Fig \ref{figure2body3body} that all the data points lie close to the same straight line, corresponding, independent of the temperature, to a slope $(p-1)=1$.

\begin{figure}
\centering
\includegraphics[width= 2.8 in]{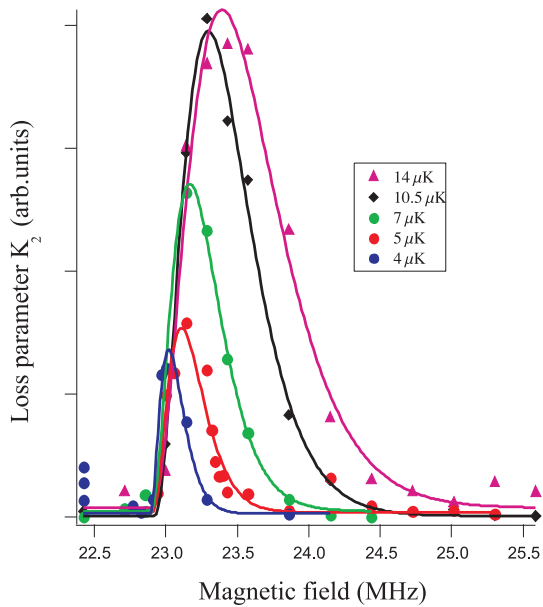}	
\caption{\setlength{\baselineskip}{6pt} {\protect\scriptsize
Loss parameter $K_2$ as a function of the magnetic field, at five different temperatures. Solid lines result from fits using eq. (\ref{eqsigmaK3int}).}} \label{lineshape}
\end{figure}

The striking conclusion of this first study is that, although two-body losses are completely excluded based on energy conservation arguments \cite{caveat}, the losses we observe, most likely due to three-body recombination, are well described by a two-body loss parameter. Such a feature will be explained in the next part of the paper. 

To go further in the analysis, we performed systematic measurements of the "two-body loss parameter" $K_2$ describing the three-body losses (thereafter called simply "loss parameter" in order to avoid confusion). 

We use two different methods to measure $K_2$. To rapidly determine $K_2$ as a function of temperature and magnetic field, we do not systematically record the atom number decay as a function of time. Rather, we use the number of atoms remaining after a given hold time $t$. Indeed, considering a fixed one-body loss parameter $\Gamma_1$, the number of atoms after a time $t$ univoquely determines $K_2$, as:

\begin{equation}
N(t)=\frac{N_0 exp(-\Gamma_1 t)}{1+\frac{K_2 n_0}{\Gamma_1 2^{3/2}}\left(1-exp(-\Gamma_1 t)\right)}
\label{Nt}
\end{equation}

The main advantage of this method is to speed up the data aquisition process, as a single data point provides a measurement of $K_2$, instead of many data points corresponding to the time-evolution of the number of atoms. Results are represented in Fig \ref{lineshape}, where (as in the rest of this paper) magnetic fields $B$ are expressed in units of frequency, $h \nu (B) = g_J \mu_B B$, where $h$ is Planck's constant, $g_J=2$ is the Land\'e factor of chromium atoms in the $^7$S$_3$ state, and $\mu_B$ the Bohr magneton. As evidenced in Fig \ref{lineshape}, we observe a reduction of both the width and the amplitude of the loss feature when the temperature decreases, in spite of the increase of the phase space density.  Below $T=2$ $\mu$K we were not able to observe any resonant loss feature around $8.2$ G. The value of the magnetic field at which losses are maximal shifts with temperature, but the value of the magnetic field above which losses start to occur does not. This value is referred to as $B_{res}$, and, as shown below, it is the value of the magnetic field at which the energy of the bound molecular level exactly coincides with the molecular dissociation limit, $i.e.$ the exact position of the Feshbach resonance.

\begin{figure}
\centering
\includegraphics[width= 2.8 in]{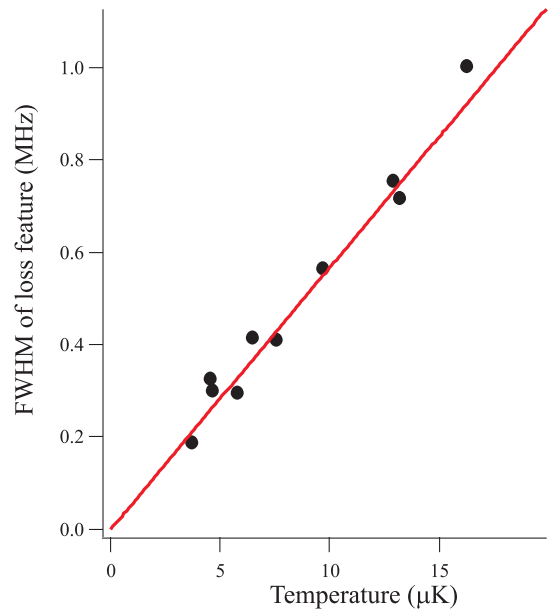}
\caption{\setlength{\baselineskip}{6pt} {\protect\scriptsize Full Width Half Maximum (FWHM) of the loss parameter feature (see Fig \ref{lineshape}), as a function of temperature. The solid line results from a fit of slope 1 to the experimental data. 
}} \label{figwidth}
\end{figure}

We also represent in Fig \ref{figwidth} the variation of the width of the observed loss feature as a function of the temperature. We find that this width increases linearly with the temperature.

\begin{figure}
\centering
\includegraphics[width= 2.8 in]{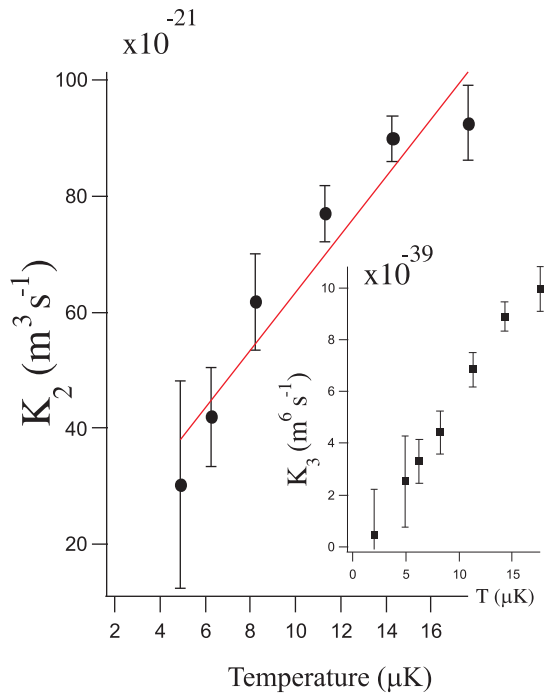}
\caption{\setlength{\baselineskip}{6pt} {\protect\scriptsize Experimentally determined two-body loss parameter.  The solid line results from a fit of slope 1 to the experimental data.  Inset: Corresponding loss parameter assuming a three-body loss parameter $\dot{n}=-K_3 \times n^3$.
}} \label{figK3T}
\end{figure}

For a precise determination of the magnitude of $K_2$, we use a different approach: we systematically record the number of atoms as a function of time, and fit the data with eq.(\ref{Nt}). We report in Fig \ref{figK3T} the temperature variation of the loss parameter $K_2^{max}$ measured, for each temperature, at the magnetic field at which losses are maximal. We find that within experimental uncertainty $K_2^{max}$ is proportional to T. This contrasts with the situation of broad s-wave Feshbach resonances, for which the loss parameter does not depend on temperature.

Let us now summarize the main experimental features of the losses close to the d-wave Feshbach resonance: (i) three-body losses are described by a two-body loss parameter $K_2$ (Fig \ref{figure2body3body}); (ii) the lineshape of the loss feature is asymmetric (Fig \ref{lineshape}); (iii) the width of the loss features increases linearly with T (Fig \ref{figwidth});  (iv) $K_2^{max}$ varies linearly with T (Fig \ref{figK3T}). To account for these features, we have developped the theoretical model described in the next part of this paper.

\section{III Theoretical interpretation}

We interpret our experimental results using the theoretical framework developed in \cite{Mies,yurovsky} for $s$ wave Feshbach resonances. Following the argumentation of these references, we consider that three-body losses happen in two steps. In a first step, two atoms collide in $l=2$. Tunneling to short internuclear distances is dramatically increased by the presence of a molecular bound state resonant with the collisional energy, and coupled to the incoming pair of atoms. Such coupling is described by the following reaction: 

\begin{equation}
Cr+Cr \longleftrightarrow Cr_2^{*} 
\label{crcr}
\end{equation}
This Feshbach resonance increases the probability of presence of two atoms at short interatomic distance. A third atomic body collides with the pair of atoms $Cr_2^{*}$ before it tunnels back out, which triggers exoergic collisions producing one more deeply bound molecule $Cr_2^{deep}$: 

\begin{equation}
Cr_2^{*}+Cr\longrightarrow Cr_2^{deep}+Cr
\label{crcr2}
\end{equation}
In the end, three chromium atoms (one dimer and one atom) have disappeared from the trap. In this framework, three-body losses can be described by the following Breit-Wigner cross-section:

\begin{equation}
\sigma(k) = \frac{\pi}{k^2} \frac{\Gamma_m (\epsilon) \Gamma_d (n)}{(\epsilon-\epsilon_0)^2/\hbar^2+(\Gamma_m (\epsilon) + \Gamma_d (n))^2/4} \equiv  \frac{\pi}{k^2} \left| S(\epsilon,n) \right|^2
\label{eqsigma}
\end{equation}
Here, $k$ is the collision wavevector, $\epsilon$ the collision energy, and $\Gamma_d (n) = \beta_d n$ is the inverse life time associated with the inelastic collision with the third body, $i.e.$ the relaxation width of the bound state $Cr_2^{*}$: $\beta_d$ is the loss parameter associated with reaction (\ref{crcr2}). $(\epsilon-\epsilon_0)$ is the energy separation to the Feshbach resonance: $\epsilon_0= \Delta M_S g_J \mu_B (B-B_{res})$, where $\Delta M_S =1$ is the difference in magnetic number between the incoming channel and the molecular bound state, is the energy of the bound state relative to the dissociation limit of the incoming channel.  Finally, $\hbar \Gamma_m (\epsilon)$ describes the strength of the resonant coupling  associated with reaction (\ref{crcr}) at a collision energy $\epsilon$. It is interesting to underline the similarity of eq. (\ref{eqsigma}), used for the study of s-wave Feshbach resonances, to the standard expression for photo-association trap loss \cite{Napolitano}, which was also applied to collisions in higher partial waves. In the Wigner-threshold regime, when the collision energy is small, $\hbar \Gamma_m (\epsilon) = A_l \epsilon^{(2l+1)/2}$ \cite{Julienne}. As a consequence, one expects that the loss parameter will depend on the incoming partial wave $l$.  

From eq. (\ref{eqsigma}), one can deduce the loss parameters using 

\begin{equation}
\dot{n}=-3 \left(n \sigma(k)v_r \right) n = -K_3(k) n^3
\label{eqsigmaK3n}
\end{equation}
with $v_r$ the thermal relative velocity. The experiment does not directly probe $K_3(k)$, but its averaged value $K_3(T)$ over the thermal distribution. If we assume a Maxwellian distribution at temperature T (see for example \cite{Napolitano}), 

\begin{equation}
K_3(T) = \frac{3}{n h Q_T} \int \left| S(\epsilon,n) \right|^2 exp(\frac{-\epsilon}{k_B T}) d \epsilon
\label{eqsigmaK3}
\end{equation}
where $Q_T=(2 \pi \mu k_B T / h^2)^{3/2}$ is the translational partition function, with $\mu=m/2$ is the reduced mass. Compared to what is shown in \cite{Napolitano}, eq. (\ref{eqsigmaK3}) includes a factor 3 because we describe the loss of three particles, and $K_3$ is reduced by a factor of $2l+1$ because only atoms in $\left|l=2,m_l=1\right\rangle$ contribute to losses. 

Very different results are obtained depending on the typical relative values of $\hbar \Gamma_m (\epsilon)$, $\hbar \Gamma_d (n)$, and $k_B T$. The goal of this paragraph is to pin-point which approximation holds in our case; consistency will be checked below. $\hbar \Gamma_m (\epsilon)$ depends on the collision energy $\epsilon$, but in the following, we will assume that the Feshbach width can be characterized by the width $\hbar \Gamma_m^{max} =  \hbar \Gamma_m(\epsilon_0^{max})$ with $\epsilon_0^{max} = \left(l+1/2\right) \times k_B T$. The reason for this choice of $\epsilon_0^{max}$ will appear later. If $\Gamma_m^{max} >> \Gamma_d (n)$, $\Gamma_d (n)$ can be neglected in the denominator of eq. (\ref{eqsigma}), and the loss rate $\dot{n}$ is proportional to $n^3$: $K_3(T)$ is then independent of $n$ and atoms losses are proportional to $N^3$. In order to reproduce our observations reported in Fig \ref{figure2body3body}, we will therefore assume in the calculation that $\Gamma_m^{max} << \Gamma_d (n)$. Furthermore, if $\hbar \Gamma = \hbar (\Gamma_m^{max} + \Gamma_d (n))$ is large compared to $k_B T$, the width of the experimental loss feature should not depend on temperature (and the three-body loss parameter should be proportional to $T^l$). This result applies for example to broad $s$ wave Feshbach resonances. When $\hbar \Gamma$ is on the same order of magnitude as $k_B T$, the experimental width should increase with $T$, but not linearly. When $\hbar \Gamma$ is very small, the lorentzian function in eq. (\ref{eqsigmaK3}) qualitatively acts as a Dirac function on the energy averaging function, and the experimental width varies linearly with $T$. Given the fact that the experimental width grows linearly with T, we therefore assume in the calculation that $(\hbar \Gamma_m^{max} << \hbar \Gamma_d (n)) << k_B T$ . Within these approximations, there is an analytical expression for eq. (\ref{eqsigmaK3}): 

\begin{equation}
K_3(T) = \frac{3}{n Q_T} \Gamma_m (\epsilon_0) exp(-\epsilon_0 / k_B T)  
\label{eqsigmaK3int}
\end{equation}

For physical interpretation, it is interesting to explicitly write eq. (\ref{eqsigmaK3n}), using $K_3(T)$ as determined by eq. (\ref{eqsigmaK3int}). Then one simply finds for $\epsilon_0 > 0$:

\begin{equation}
\dot{n}=-\alpha \left( n \Lambda_{dB}^3\right) \Gamma_m (\epsilon_0) exp(-\epsilon_0 / k_B T) \times n  
\label{eqpsd}
\end{equation}
where $\Lambda_{dB} = \frac{h}{\sqrt{2 \pi m k_B T}}$ is the thermal de Broglie wavelength, and $\alpha = 6 \sqrt{2}$ is a numerical factor. Maximum losses occur for $\epsilon_0^{max}$. We note that we explicitly find in eq. (\ref{eqpsd}) that the rate of association is, in our case, explicitly determined by the phase-space density, as also observed when the Feshbach resonance is wide \cite{Kohler}. 

We can now give a simple physical interpretation of eq. (\ref{eqpsd}). We note $\left|\psi _{bound}\right\rangle$ the molecular bound state wavevector,  $\left|\psi_{\epsilon}\right\rangle$ the incoming channel wavevector, and $H_{dd}$ the dipole operator coupling these two states. The Feshbach coupling is therefore defined as $\Gamma_m(\epsilon) = 2\pi \left| \left\langle \psi _{bound} | H_{dd} | \psi_{\epsilon} \right\rangle \right|^2$, which can also be written, in a trap, as $2 \pi V^2 \rho (\epsilon) = \hbar \Gamma_m(\epsilon)$ where $\rho(\epsilon)=\frac{dv}{d \epsilon}$ is the density of states in the trap ($v$ denotes the trap vibrational states) and $V$ is the matrix element for coupling between the bound state and one vibrational trap state \cite{Mies}. If we rewrite the expression of the density of states as $\rho(\epsilon) = \frac{\Delta v}{\Gamma_d}$, where $\Delta v$ is the number of vibrational states in the trap covering an energy width set by $\Gamma_d$, then  $N^{res} = \Delta v  \left( n \Lambda_{dB}^3\right)$ is the number of atoms in the trap which are resonant with the bound molecular state of width $\Gamma_d$. Therefore, eq. (\ref{eqpsd}) leads to an order of magnitude of the maximum loss rate:
\begin{equation}
\frac{\dot{n}}{n} \propto \frac{V^2}{\hbar^2 \Gamma_d} \times N^{res}   
\label{Fermigolden}
\end{equation}
We thus recover a Fermi golden rule: each trap vibrational level is an independant incoming channel weakly coupled to a state whose lifetime is so short that it behaves like a continuum whose density of states is $\frac{1}{\Gamma_d}$ \cite{thanks}. 

\section{IV Quantitative comparison with experimental data}

We now verify that eq. (\ref{eqsigmaK3int}) quantitatively reproduces all the experimental findings described above. First, we find that $K_3$ is proportional to $1/n$: the experimental loss rate is proportional to $n^2$, not to $n^3$, although three particles are indeed needed for the loss to occur. In fact, $\Gamma_m^{max} << \Gamma_d (n)$ corresponds to a situation where the collisionnally limited lifetime of the molecules due to reaction (\ref{crcr2}) is much shorter than the timescale for this molecule to be coupled back to the continuum through Feshbach coupling. For this reason, the experimental loss rate is only set by the slow reaction (\ref{crcr}), which is a two-body process. This is why $\Gamma_d$ does not appear in eq. (\ref{eqsigmaK3int}). Only in the limit of very small densities should we recover a loss rate proportional to $n^3$. This result is in agreement with our experimental findings evidenced in Fig \ref{figure2body3body}: three-body losses are in the present case best described by a two-body loss parameter $K_2=K_3 \times n$. This is a signature that indeed $\Gamma_m^{max} << \Gamma_d (n)$: the precise study of the dynamics of the atom number decay therefore provides an indirect signature of the short lifetime of the molecular bound state. Note that in the inset of Fig \ref{figK3T}, we also report a three-body loss parameter to enable the comparison with three-body loss parameters in other systems. 

We also find that the experimental lineshapes (shown for example in Fig \ref{lineshape}) are well fitted by eq. (\ref{eqsigmaK3int}). Temperature is used as a free parameter in these fits, and this yields fitted temperatures close to the ones experimentally determined by analysing expanding clouds in free fall, at any temperature between 3 $\mu$K and 15 $\mu$K. Note that the width of the theoretical lineshape is proportional to the temperature, consistent with our observations reported in Fig \ref{figwidth}. The temperatures given by the fit are nevertheless systematically smaller than the measured temperatures by about 20 percent. Such discrepancy may arise from the fact that, when the loss rate is high ($\epsilon_0 \approx \epsilon_0^{max}$), there is a slight increase in temperature, which in turn raises the loss parameter. This effect may decrease the half-width at half maximum of the loss feature that we measure. Another possibility that we cannot exclude is that our temperature measurements are slightly distorted by the presence of a small stray magnetic field with spatial curvature. 

We now turn to the temperature dependence of the loss parameter. As shown in Fig \ref{figK3T}, the maximum loss rate parameter $K_2^{max}$ increases linearly with temperature. This is in agreement with eq. (\ref{eqpsd}) which shows that the maximal theoretical two-body loss parameter ($i.e.$ at $\epsilon_0^{max}$) varies as $T^{l-1}$, therefore as $T$ for a d-wave Feshbach resonance.  

Thus, eq. (\ref{eqsigmaK3int}) reproduces all the experimental features with a very satisfying accuracy. We can therefore in turn use it to deepen the knowledge on this d-wave Feshbach resonance. For example, from the analysis of the loss parameter, we can measure precisely the position of the Feshbach resonance. This position is not the magnetic field at which losses are maximal, but rather, the position at which losses start to occur, which precisely happens when the bound molecular state crosses the open channel asymptote. Our measurements indicate: $B_{res}= (8.155 \pm 0.015) G$. Fields are precisely calibrated in the vicinity of the Feshbach resonance by rf atomic spectroscopy. The precision in our measurement is a factor of 6 better than what was previously reported \cite{stuhler}. 

Comparing our experimental result to eq. (\ref{eqsigmaK3int}), we also deduce that $\frac{\hbar \Gamma_m^{max}}{k_B T} \approx (5.5 \pm 1.2 \pm 2.5) \times 10^{-5} \left(\frac{k_B T}{8 \mu K}\right)^{3/2}$. The error bars represent respectively the statistic error bar, and the systematic error bar, dominated by the uncertainty on the oscillation frequencies of the trap. For example, at a temperature of 8 $\mu$ K, $\Gamma_m^{max}/2 \pi \approx 10$ Hz. It is striking that, although the Feshbach coupling is so small that at any given time only a small fraction ($\approx$ 10$^{-4}$) of the atoms can experience it, the Feshbach resonance can nevertheless lead to substantial losses, if one waits long enough. Thermalization constantly insures that there are atoms whose collision energy matches $\epsilon_0^{max}$. 

Having determined $\Gamma_m^{max}$, we can now check for the consistency of the approximations which lead to eq. (\ref{eqsigmaK3int}). Clearly $\hbar \Gamma_m^{max} << k_B T$. The exact timescale for vibrational relaxation is not known in the case of Cr, but it is expected to be on the same order of magnitude as for other bosonic species, i.e. $\Gamma_d (n) / 2 \pi \approx 1 $ kHz. The approximation $\hbar \Gamma_m^{max} << \hbar \Gamma_d (n) << k_B T$ is therefore most likely valid. More precisely, $\hbar \Gamma_d (n) << k_B T$ is obvious at the experimental densities, while  $\Gamma_m^{max} << \Gamma_d (n) $ implies molecular lifetimes of less than 10 ms. The good agreement between our theoretical model and our measurements thus indirectly provides an upper-bound for the collisionnally limited molecular lifetime of the bound state. 

There are three main conclusions to this study. First, our analysis allows us to precisely measure the position of the resonance. Second, the width of the Feshbach resonance $\Gamma_m$ has been measured, and it is very small. This arises from the fact that, in a $d$ wave Feshbach resonance, particles need to tunnel through a centrifugal barrier to reach the molecular bound state. We also show that $\Gamma_m << \Gamma_d (n)$, which indicates that it should be difficult to create a large number of cold molecules by sweeping a magnetic field through the Feshbach resonance, as achieved in other experiments using a $s$ wave Feshbach resonance \cite{Kohler}. A third information is that $K_2 \propto T$; going to lower and lower temperatures insures that losses can be strongly reduced. 

Eq. (\ref{eqpsd}) also shows that, in the regime where the collisional lifetime of the Feshbach molecules $1/\Gamma_d$ is short, it is not likely to use such d-wave Feshbach resonances in a BEC to tailor anisotropic interactions: for the interactions to have a substantial effect on the trapped atoms, one needs to have $\Gamma_m^{max}$ larger than the trapping frequencies, but then the lifetime of the cloud is shorter than the trap period when $\left( n \Lambda_{dB}^3\right) \approx 1 $.

\section{V Theoretical estimate of the Feshbach coupling}

Losses as predicted by Eq. (\ref{eqpsd}) depend on a single parameter $\Gamma_m(\epsilon_0)$. We now turn to the comparison between the measured value of $\Gamma_m(\epsilon_0)$ and a theoretical estimate obtained by calculating the coupling between the collisional channel and the bound molecular state by dipole-dipole interactions. As stated above, the parameter $\Gamma_m (\epsilon)$ is defined as $2\pi |< \psi _{bound} | H_{dd} | \psi_{\epsilon} >|^2$, and depends on the matrix element of the dipole-dipole interaction between the bound molecular state and the d-wave collisional state with energy $\epsilon$. We recall the expression of the dipole-dipole operator
%%%%%%%%%%%%%%%%%%%%%%%%%%%%%%
\begin{eqnarray}
\label{Hdd}
H_{dd}=4\mu_B^2 \frac{\mu_0}{4\pi} \frac{(\vec{S}_1.\vec{S}_2)-3(\vec{S}_1.\hat{R})(\vec{S}_2.\hat{R})}{R^3},
\end{eqnarray}
%%%%%%%%%%%%%%%%%%%%%%%%%%%%%%
where $\vec{S}_1$ and $\vec{S}_2$ are the spin operators of the two interacting atoms with interatomic distance $R$ along the internuclear axis $\hat{R}$, $\vec{R}=R\hat{R}$.
In tensor operator description, one has
%%%%%%%%%%%%%%%%%%%%%%%%%%%%%%
\begin{eqnarray}
\label{Hdd-tensoriel}
H_{dd}=-4\sqrt{6}\mu_B^2 \frac{\mu_0}{4\pi} \frac{\big \{ \vec{S}_1.\vec{S}_2\big \}^2_0}{R^3},
\end{eqnarray}
%%%%%%%%%%%%%%%%%%%%%%%%%%%%%%
where the projection of the tensor rank is taken on the internuclear axis. 
By separating angular and radial parts, as allowed by the Born-Oppenheimer approximation, the relevant states of an atomic pair can be written as 
%%%%%%%%%%%%%%%%%%%%%%%%%%%%%%
\begin{eqnarray}
\label{cont}
|1> = |S=6,~M_S=-6,~\ell=2,~m_{\ell}=1>~F_{1} (R)
\end{eqnarray}
%%%%%%%%%%%%%%%%%%%%%%%%%%%%%%
for the collision channel, with $F_1(R)$ the energy normalized radial wavefunction, and
%%%%%%%%%%%%%%%%%%%%%%%%%%%%%%
\begin{eqnarray}
\label{lie}
|2> = |S=6,~M_S=-5,~\ell=0,~m_{\ell}=0>~F_{2} (R)
\end{eqnarray}
%%%%%%%%%%%%%%%%%%%%%%%%%%%%%%
for the bound molecular state, which is the first bound level starting from the dissociation limit. $F_2(R)$ is the radial wavefunction of this molecular state. The angular part of the matrix element of eq. (\ref{Hdd-tensoriel}) is easily evaluated using standard tensor operator technique, remembering that the angular momentum projections appearing in eqs (\ref{cont},\ref{lie}) are taken on a fixed axis (determined by the static magnetic field):

\begin{eqnarray}
<(S_1~S_2)~S~M_S~\ell~m_{\ell}|\big \{ \mathbf{S}_1.\mathbf{S}_2\big \}^2_0 |(S_1~S_2)~S'~M'_S~\ell'~m'_{\ell}>=\nonumber \\
\delta (M_S+m_{\ell},M'_S+m'_{\ell}) (-1)^{S-M_S}\nonumber \\
\sqrt{5(2S+1)(2S'+1)(2\ell +1) (2\ell '+1)} \nonumber \\
\Big{[}S_1(S_1+1)(2S_1+1)S_2(S_2+1)(2S_2+1) \Big{]}^{1/2} \nonumber \\
\left(
\begin{array}{ccc}
\ell & 2 & \ell' \\
0 & 0 & 0 \\
\end{array}
\right)
\left\{
\begin{array}{ccc}
S_1 & S_1 & 1 \\
S_2 & S_2 & 1 \\
S & S' & 2 \\
\end{array}
\right\} \nonumber \\
\sum_{p=-2,2}{(-1)^{m'_{\ell}+p}}
\left(
\begin{array}{ccc}
\ell & 2 & \ell' \\
-m_{\ell} & -p & m'_{\ell} \\
\end{array}
\right)
\left(
\begin{array}{ccc}
S & 2 & S' \\
-M_S & p & M'_S \\
\end{array}
\right)
\end{eqnarray}

The radial wavefunctions are described in a simple, purely asymptotic model. The method is based on the concept of nodal lines \cite{crubellier99,vanhaecke04}. For energy values close to the dissociation limit (either above or below), the nodal structure of the radial wavefunctions in the inner part of the potential varies very slowly with energy. In a first approximation, the radial position of the nodes of the wavefunctions are located on straight lines as a function of collision energy, which characterize entirely the inner part of the potential, which can thus be ignored. The radial Shr\" odinger equations are solved in the asymptotic part only ($i.e.$ at large interatomic distances), with the constraint that the wavefunction vanishes on a given nodal line, chosen in the inner/asymptotic frontier region. The nodal lines have to be adjusted to reproduce experimental results (scattering length values, Feshbach resonances, near-threshold bound level positions, etc.). We use here a simplified version of the method, which consists of a $R^{-6}$ potential limited at short range by an infinite repulsive wall, having a tuneable position chosen to reproduce the scattering length of the system. The model depends thus on two parameters, the van der Waals $C_6$ coefficient and the position of the wall, i.e. of the chosen nodal line \cite{crubellier06}.  The node position is adjusted to roughly reproduce the energy of the first bound level of the $S=6$ molecular potential of Cr$_2$, -23MHz, which is deduced from the measurements of Feshbach resonances of ref. \cite{werner} and corresponds to a scattering length of 103~a$_0$. The $C_6$ value (733~atomic units) is also taken from \cite{werner}.

The calculation of the free and bound radial wavefunctions is performed by inward numerical integration of the radial differential equations, starting from large $R$ values. For the bound state, asymptotic exponentially decreasing behavior is imposed and the energy is found iteratively to ensure that the solution vanishes at the node position. For the continuum, two solutions are calculated, with respective asymptotic behavior $\sin(kx)$ and $\cos(kx)$; the wavefunction is the linear combination of these solutions vanishing at the node position. The square of the radial integrals
%%%%%%%%%%%%%%%%%%%%%%%%%%%%%%
\begin{eqnarray}
\label{rad-int}
I(\epsilon)=\int_0^\infty{\frac{F_{1} (R)F_{2} (R)}{R^3}} R^2 dR
\end{eqnarray}
%%%%%%%%%%%%%%%%%%%%%%%%%%%%%%
gives the energy dependence of $\Gamma _m(\epsilon)$, which is found to obey the Wigner law 
%%%%%%%%%%%%%%%%%%%%%%%%%%%%%%
\begin{eqnarray}
\label{wigner}
\Gamma_m(\epsilon)\propto \epsilon ^{5/2}
\end{eqnarray}
%%%%%%%%%%%%%%%%%%%%%%%%%%%%%%
in a large energy range (the relative discrepancy is less than 1\% at 100$\mu$ K). For $k_BT=8 \mu$ K, the calculation gives 
%%%%%%%%%%%%%%%%%%%%%%%%%%%%%%
\begin{eqnarray}
\label{gamamax}
\frac{\Gamma_m^{max}}{k_BT}= (7.3 \pm 0.3)\times 10^{-5},
\end{eqnarray}
%%%%%%%%%%%%%%%%%%%%%%%%%%%%%%
which is very close to the experimental value, $( 5.5 \pm 1.2 \pm 2.5) \times 10^{-5}$.

\section{VI Conclusion}

In this paper, we have analysed a Feshbach resonance resonantly increasing three-body losses associated to collisions in $d-$wave. As the atoms have to tunnel through a centrifugal barrier before experiencing the resonant bound state, the Feshbach coupling is small compared to the inverse lifetime of the bound molecular state. As a consequence:
\begin{itemize}

\item atom losses due to three-body recombination are described by a two-body loss parameter $K_2$;

\item as the width of the Feshbach coupling is small compared to $k_B T$, the observed resonance width increases linearly with $T$;

\item the loss parameter peak value (as a function of the magnetic field) increases linearly with $T$.

\end{itemize}

All these features are well accounted for by a theoretical model with no free parameter. We directly relate the amplitude of the losses to the coupling between the incoming channel and a molecular bound state, due to dipole-dipole interactions. In addition, our measurements provide a better determination of the position of this d-wave Feshbach resonance, in addition to an upper bound to the lifetime of the molecular bound state.

Acknowledgements: LPL is Unit\'e Mixte (UMR 7538) of CNRS and of Universit\'e Paris Nord. We acknowledge financial support from Minist\`{e}re de l'Enseignement Sup\'{e}rieur et de la Recherche (CPER), IFRAF (Institut Francilien de Recherche sur les Atomes Froids) and from the Plan-Pluri Formation (PPF) devoted to the manipulation of cold atoms by powerful lasers.


\begin{thebibliography}{99}

\bibitem{tunable} M. Gustavsson et al Phys. Rev. Lett., \textbf{100}, 080404 (2008), M. Fattori et al. Phys. Rev. Lett., \textbf{100}, 080405 (2008)

\bibitem{Kohler} T. Kohler, K. Goral, and P. Julienne, Rev. Mod. Phys., \textbf{78}, 1311 (2006)

\bibitem{bloch} I. Bloch, J. Dalibard, and W. Zwerger, Rev. Mod. Phys., \textbf{80}, 885 (2008)  

\bibitem{regal} C. A. Regal, C. Ticknor, J. L. Bohn and D. S. Jin, Phys. Rev. Lett., \textbf{90}, 053201 (2003) 

\bibitem{buggle} Ch. Buggle, J. L\'eonard, W. von Klitzing, and J. T. Walraven, Phys. Rev. Lett., \textbf{93}, 173202 (2004)

\bibitem{sweeprf} Q. Beaufils et al., Phys. Rev. A \textbf{77}, 053413 (2008)

\bibitem{caveat} Losses due to two-body elastic collisions followed by evaporation from the finite depth optical trap cannot be completely excluded. We have checked that they are negligible away from the Feshbach resonance.  

\bibitem{werner} J. Werner et al., Phys. Rev. Lett. \textbf{94}, 183201 (2005) 

\bibitem{beaufilsBEC} Q. Beaufils et al., Phys. Rev. A \textbf{77}, 061601 (2008)

\bibitem{Mies} F. H Mies, E. Tiensinga, and P. S. Julienne, Phys. Rev. A, \textbf{61}, 022721 (2000).

\bibitem{yurovsky} V. A. Yurovsky and A. Ben-Reuven, Phys. Rev. A, \textbf{67}, 050701(R) (2003). 

\bibitem{Napolitano} R. Napolitano et al., Phys. Rev. Lett., \textbf{73}, 1352 (1994).

\bibitem{Julienne} P. S. Julienne and F. H. Mies, J. Opt. Soc. Am. B. \textbf{6}, 2257 (1989).

\bibitem{thanks} We thank W. D. Phillips for pointing this physical interpretation to us.

\bibitem{stuhler} J. Stuhler et al., J. Mod Opt. \textbf{54}, 647 (2007)

\bibitem{crubellier99} A. Crubellier et al., Eur Phys. J. D., \textbf{6}, 211 (1999)

\bibitem{vanhaecke04} N. Vanhaecke et al., Eur Phys. J. D., \textbf{28}, 351 (2004)

\bibitem{crubellier06} A. Crubellier and E. Luc-Koenig, J. Phys. B., \textbf{39}, 1417 (2006)

\end{thebibliography}
\end{document}